# A fully woven touchpad sensor based on soft capacitor fibers

Jian Feng Gu, Stephan Gorgutsa, Maksim Skorobogatiy

Génie Physique, École Polytechnique de Montréal, Montréal H3C 3A7, Canada

E-mail : maksim.skorobogatiy@polymtl.ca

Abstract: A novel, highly flexible capacitor fiber (with 100 nF m$^{-1}$ typical capacitance per length) having a multilayer periodic structure of dielectric and conductive polymer composite films is fabricated by drawing technique. The fiber is used to build a touchpad sensor. Then, we study the influence of the fiber length, capacitance and volume resistivity on the touch sensing performance. A theoretical ladder network model of a fiber network is developed. A fully woven textile sample incorporating one-dimension array of the capacitor fibers is fabricated. Finally we show that such an array functions as a two-dimensional touch sensor.

1. Introduction

Touch sensing as a human interface device (HID) technology, for example to replace the computer mouse, is becoming increasingly popular. Various sensing systems can be developed based on the different principles including resistive, capacitive, infrared, surface acoustic wave, electromagnetic, near field imaging, etc. Resistive [1] and capacitive [2] methods have been widely used in conventional touch screens of commercial products such as mobile phones, PDAs, and consumer electronics devices. Resistive touch screens are composed of two sheets that are coated with a resistive material, commonly indium tin oxide (ITO), and separated by an air gap or microdots. When a finger presses the screen, the two sheets connected at the touch position which changes the current flows on the screen. A sensing circuit can detects the changes and located the touch. A capacitive touch sensor is based on capacitive coupling effects. A typical design is to coat screen with a thin, transparent metallic layer, thus form a collection of capacitors. When a user touches the surface, the disturbance caused by the finger changes the capacitance and current that flows on the display. Infrared (IR) touch systems [3] use an array of infrared (IR) light-emitting diodes (LEDs) on two adjacent bezel edges of a display with photosensors placed on the two opposite bezel. An object, such as a finger, that touches the screen interrupts the light beams, can be detected by photosensors. The surface acoustic wave (SAW) touchscreens [4] consist of a glass panel with piezoelectric transducers, which act as ultrasonic emitters and receivers. The emitters produce ultrasound waves at the surface of the glass. When the user touches the screen, some of the acoustic waves are absorbed and detected by the receivers.

One limitation in all of these technologies is that they are only capable of detecting a single touch. Multi-touch techniques allow touch screens to recognize touches of multiple fingers or inputs of multiple persons simultaneously.  The multi-touch detection mechanisms can be classified into three categories: sensor array, capacitive sensing, and vision and optical based. A sensor array touch surface consists of a grid of touch sensors that work independently. When a user exerts multiple touches on the surface, the system can indentify activated sensors and determine these touch positions simultaneously. An example is FMTSID (Fast Multiple-Touch-Sensitive Input Device) [5], one of the first multi-point touch sensor-based devices. The system consists of a sensor matrix panel, the ranks of select register, A/D converter and a control CPU. The design of the sensor matrix is based on the technique of capacitance measurement between a finger tip and a metal plate. Another example is MTC Express produced by Tactex Controls. It features sensor array detection of pressure-sensitive touch sensors. Chen [6] used it to build a simulation of a liquid flow field in which fingers assert the obstacles. The system allows participants to interact using their whole hand with a flow of circulating particles, providing visual and auditory feedback. A capacitive touch method uses the capacitive coupling between two conductors to sense a touch. Typically, the touch surface contains a mesh of horizontal and vertical antennas which function as either a transmitter or a receiver of electric signals. Examples based on this technology include Rekimoto's SmartSkin [7], and DiamondTouch [8] developed by Dietz and Leigh. Apple has recently been granted a patent for a "multipoint touchscreen" [9] which also based on a capacitive sensing device. Vision based multi-touch technology requires a video camera to monitor the finger locations and image processing algorithms to process signals. This technology results products including DigitalDesk [10], Visual Touchpad [11], TouchLight [12], and Microsoft Surface's multiple-touch tabletop. To increase the detection accuracy and reduce the cost of the vision based system, Han [13] reported a rear-projected interactive screen by introducing the frustrated total internal reflection (FTRI) technique.  An acrylic touch surface traps the infrared light within it by total internal reflection. When an object touches the surface, it can frustrate the light reflection and scatter it at the contact point. To overcome the portability issue and improve the touch detection a ThinSight [14] optical system is implemented using a novel optical sensing technique. An array of retro-reflective optosensors is placed behind a normal LCD. The optosensors can emit IR light and detect light reflected by objects such as fingertips in front of the screen.

In addition to touchscreens, touch sensors with force sensing ability has been studied as tactile sensors for many years [15].  Such sensors have found their applications in artificial skin for robot applications [16], minimally invasive surgery [17], wearable computers [18], and mobile or desktop haptic devices [19]. Up to date tactile sensors have mainly focused on silicon-based sensors that use piezoresistive [20, 21] or capacitive sensing mechanism [22–24]. The silicon tactile sensors have limitations of mechanical brittleness, and hence are not capable of sustaining large deformations. Polymer-based tactile sensing approaches that use piezoelectric polymer films [25-27], pressure-conductive rubber [28], carbon fiber based polymer composite [29] and conductive polymers [30], have been reported as well. These sensors provide good resolution but the applied force range is low due to the limited thickness of membrane. Other tactile technologies that have been proposed in literatures include optical [31], ultrasonics-based [32], and magnetoresistive [33]. Tactile sensors based on quantum tunnel composites (QTC) have come up recently. QTC has the unique capability of transforming from a virtually perfect insulator to a metal like conductor when deformed [34].

In this article we will demonstrate that the capacitor fiber we developed recently can be used as a building block for touch sensors. We will first study the sensing performance of a single fiber and show how its behaviours are affected by different parameters. Then, we will show that these fibers can be integrated into a textile that functions as a two-dimensional touch sensor.

## 2. A Soft capacitor fiber for electronic textiles

A novel, highly flexible fiber with high electric capacitance is developed recently in our lab [35]. The fiber is based on dielectric polymers and a conductive polymer composite. It is fabricated by thermal drawing technique that consists of three steps of co-rolling multilayer films into a preform, consolidating the preform at an adequate temperature, and drawing the preform into fibers by heating in a specific furnace. After a successful drawing the resultant fibers preserve the structured profile of a preform but with a much smaller dimension scale in cross-section. Several designs of the capacitor fibers were realized according to different geometries and connectorization methods. The fiber we produced has a sub-millimeter diameter with a typical capacitance of 100 nF m$^{-1}$, which is about three orders of magnitude higher than that of a coaxial cable of comparable diameter. The capacitor fibers are ideally suited for the integration into textile products as they are soft, small diameter, lightweight and do not use liquid electrolytes.

Among the several types of designs, the fiber that will be studied in this work features a periodic sequence of conductive and isolating plastic layers positioned around metallic electrodes in its cross-section, as shown in figure 1(a). A copper wire of a diameter of 0.12mm was embedded in the fiber during the drawing process. This approach not only improves the drawability of the multi-material preforms but also ease the connectorization of the fiber into circuits. In this design the copper wire servers as one probe and the exposed outside conductive polymer layer servers as another probe. Copper wires with thickness of 0.12mm are wrapped around the fiber at the measurement points acting as electrodes to maintain a good contact between oscilloscope probe and surface of the fiber.

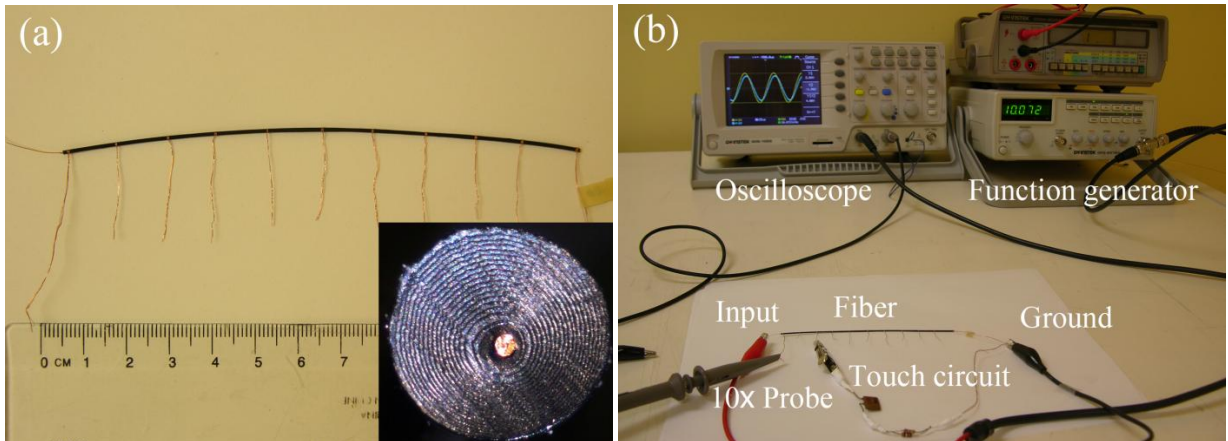

Figure 1. (a) A capacitor fiber with one cooper wire embedded in the center. Copper wires wrapped around the fiber acts as electrodes on the outer surface. Inset is the cross section view of the fiber. (b) Experimental setup of the fiber sensing measurement.

Our previous study [35] shows that the fiber capacitance is a very stable parameter independent of the fiber diameter, operational temperature and electrical probe structure. In contrast, fiber resistivity has a very strong positive temperature coefficient. It is sensitive to stretching and strongly dependent on the shape of an outer electrode. While fiber capacitance is proportional to the fiber length, fiber resistivity parameter is inversely proportional to the fiber length (assuming full electrode coverage). We have also demonstrated that due to high resistivity of conductive composite films the capacitor fiber behaves like an RC ladder circuit. At low frequencies both equivalent capacitance $C_F$ and equivalent resistance $R_F$ are constants but they decrease at frequencies higher than a particular value.

**3. The properties of a capacitor fiber as a distributed sensor**

The relative high resistivity of the conductive polymer layers endows the fiber with some distributed electrical properties. We found that when the fiber is touched by a human finger, it causes a disturbance of the distributed voltage which is related to the touch position. The purpose of this study is to find out the fiber responses to a finger touch, and the possibility to utilize this property to fabricate touch sensors.

A measurement setup is shown in Figure 1(b), where a sinusoidal signal of tuneable frequency is imposed on the fiber by connecting its inner copper wire probe to channel 1 of function generator and grounding the out surface of the fiber at the right end. The voltage of the fiber surface at different positions was acquired by an oscilloscope (GDS-1022, Good Will Instrument Co., Ltd) through a 10X probe (GTP-060A-4, Good Will Instrument Co., Ltd) on channel 2. A typical voltage distribution over a fiber is displayed on figure 2(a). Here $x/L$ denotes the measurement position normalized by fiber length, and $|V(x)/V0|$ is the voltage amplitude on the fiber surface normalized by the output voltage of function generator. When a human finger touches the fiber surface with the body grounded as shown in figure 1, a disturbance on the voltage distribution can be observed. Figure 2(b) shows the responses of this fiber when a finger touches at the position of 0.3L. We can see that at low frequencies, e.g. 10Hz and 100Hz, the finger touch does not have a noticeable effect on the voltage distribution. However at higher frequencies, e.g. 1k Hz and 10k Hz, a dip of voltage take places at the point of touch. This is because the human body is largely composed of conductive electrolytes covered by a layer of dielectric skin. Thus it can be equivalently looked as a circuit of a resistor in series connected by a capacitor. When this circuit is connected into the measurement circuit, it will affect the voltage distribution of the fiber. In this study we used a 1.44kΩ resistor and a 0.15nF capacitor as equivalent circuit to replace the finger touch. Our experiments confirm that this circuit has a similar effect as the finger touch, but with a better repeatability. So all the data of finger touch presented in the following paragraphs, if not mentioned specifically, are all acquired by the equivalent touch circuit shown in figure1(b).

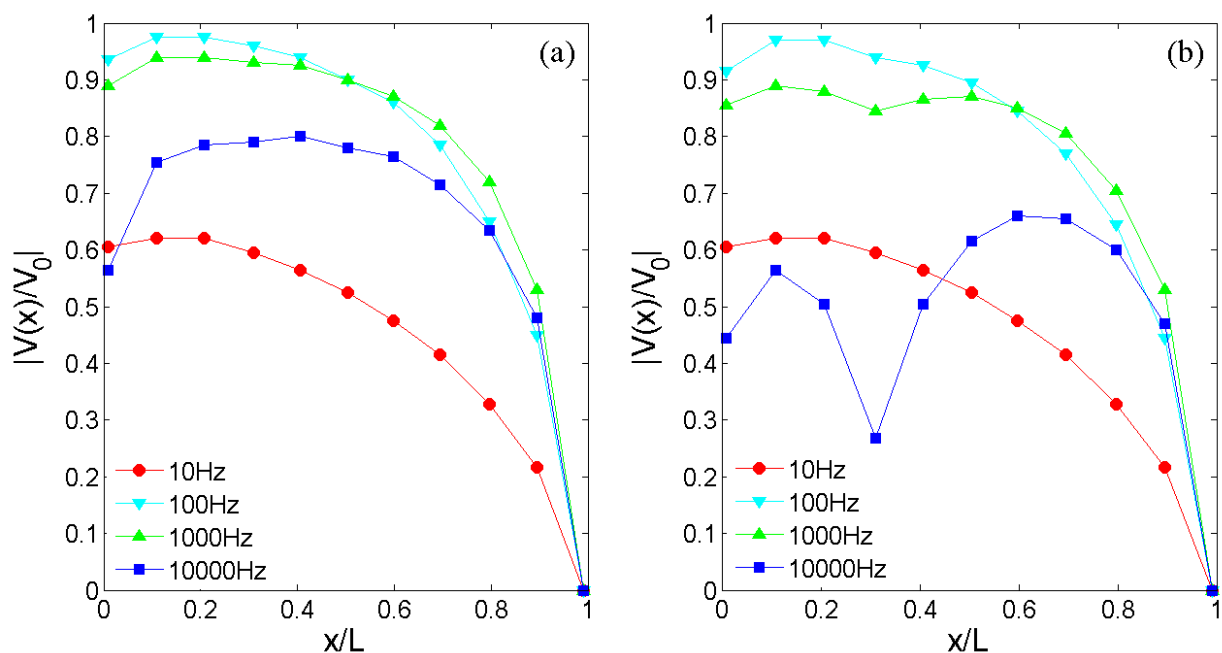

Figure 2. Comparison of voltage distribution for a fiber without (a) and with (b) a finger touch.

A finger touch can affect the voltage distribution of a fiber. Inversely, such voltage disturbance can also be used as a signal to sense the finger touch. Moreover, we can even determine the exact position of the touch through some accurate voltage measurements. For example, we can place an array of electrodes on the fiber surface along the length and monitor the position of voltage dip to determine the touch position. An alternative easier way is to record the voltage at the left end of the fiber as a function of touch position, as shown in figure 2(c). Here $x_b/L$ represents the touch position of the finger, normalized by fiber length. $|V(0)/V_N(0)|$ represents the voltage at the left end of the fiber, i.e. x=0, and is normalized by the value without finger touch. We can see that at low frequencies (10 and 100Hz), $|V(0)/V_N(0)|$ is almost a constant indepedant on the touch position, while at high frequencies (1k and 10kHz), there exists a monotonous relationship between them over most the part of fiber. This one-to-one mapping enables us to determine touch position by the value of voltage monitored at the end of a fiber. Figure 2(c) demonstrates that the fiber has a better resolution at a higher frequency of 10kHz because of its wider range of relative voltage variation, while it cannot detect the finger touch as frequency decreases to 100 Hz.

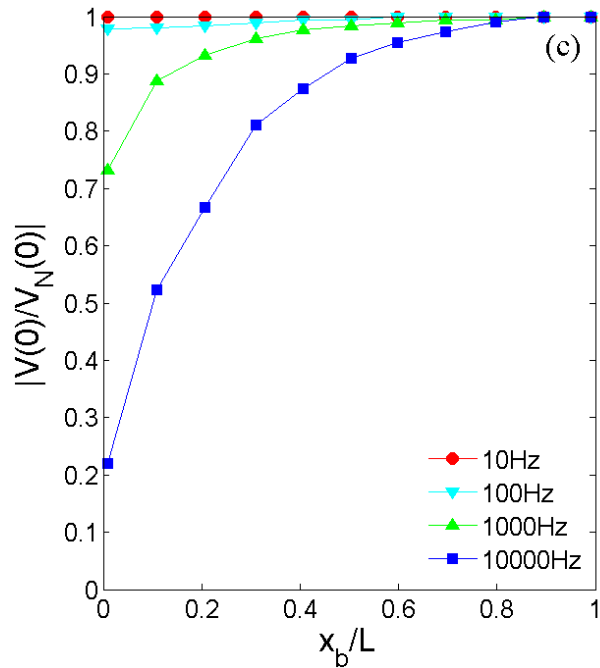

Figure 2(c). The voltage at x=0 as a function of touch position

## 4. Theoretical model

In order to analyze the above observations, we developed a theoretical model for the sensing performance of an individual fiber. This model will provide us design guidelines for a sensing textile. To do this we depict the capacitor fiber as a ladder network with infinite number of small resistance and capacitance (RC) elements as shown in Figure 3. Here $dR(\omega)$ and $dC(\omega)$ are the equivalent transverse resistance and capacitance of a fiber segment of a length of $dx$. The touch is take place at the position of $x_b$. $R_p$ and $C_p$ are the equivalent resistance and capacitance of the10X probe and oscilloscope. We have to take the input impedance of the oscilloscope into consideration, because its value is comparable to the fiber impedance and cannot be neglected in the measurement circuit.

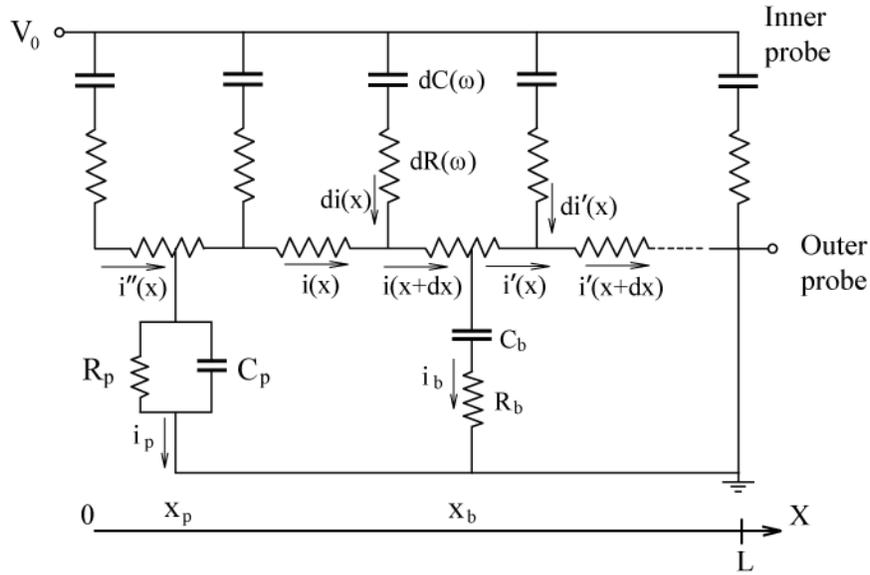

Figure 3. The ladder network model of a fiber touched by a finger

Our previous study [35] gives the following expressions for the transverse resistance and capacitance of a fiber:

$$dC(\omega) = -\frac{dx}{\omega r_t \cdot Im(f(B))} \quad (1)$$

$$dR(\omega) = \left(\frac{1}{2} + Re(f(B))\right)\frac{r_t}{dx} \quad (2)$$

where

$$r_t \approx \frac{S\rho_v}{d_c}, \quad r_l \approx \frac{\rho_v}{Sd_c},$$

$$f(B) = \frac{1+\cosh(B)}{B \cdot \sinh(B)}; B = \sqrt{2j\omega R^t C}, C \approx 2\varepsilon_0\varepsilon\frac{S}{d_i}$$

Here $\rho_v$ is the volume resistivity of conductive films, $L$ is the length of the fiber, $S$ and $d_c$ are respectively the width and thickness of the conductive films. $\varepsilon$ is dielectric constant of the isolating films, $\varepsilon_0$ is permeability of the vacuum, $d_i$ is thickness of the rolled isolating films. $i''(x)$ denotes the current flowing in the conductive film before the oscilloscope probe, $i(x)$ denotes the current between the probe and touch position, and $i'(x)$ is the current after the touch point. $V_0$ is the voltage difference between the inner probe and outer probe at $x=L$. We assume that the resistivity of the conductive film is a position-independent and frequency-independent parameter. If the measured position is on the left side of the touch point, i.e., $x_p<x_b$, as shown in figure 1, applying KVL and KCL to the ladder circuit leads to the following equations:

$$di(x)\left(\frac{1}{j\omega dC} + dR\right) + \int_x^{x_b} r_l i(l)dl = V_0 - V(x_b) \quad (3)$$

$$di'(x)\left(\frac{1}{j\omega dC} + dR\right) + \int_x^L r_l i'(l)dl = V_0 \qquad (4a)$$

$$di''(x)\left(\frac{1}{j\omega dC} + dR\right) + \int_x^{x_p} r_l i''(l)dl = V_0 - V(x_p) \qquad (5a)$$

$$i_b\left(\frac{1}{j\omega C_b} + R_b\right) = V(x_b) \qquad (6)$$

$$i_p \frac{R_p}{1 + j\omega C_p R_p} = V(x_p) \qquad (7)$$

The boundary conditions are
$i''(0) = 0$, $i(x_b) = i'(x_b) + i_b$ and $i''(x_p) = i(x_p) + i_p$ \qquad (8a)

If the measured position is on the right side of the touch point, i.e., $x_p > x_b$, equations (4a) and (5a) are replaced by (4b) and (5b) respectively,

$$di'(x)\left(\frac{1}{j\omega dC} + dR\right) + \int_x^L r_l i'(l)dl = V_0 - V(x_p) \qquad (4b)$$

$$di''(x)\left(\frac{1}{j\omega dC} + dR\right) + \int_x^{x_p} r_l i''(l)dl = V_0 \qquad (5b)$$

And the boundary conditions become:
$i(0) = 0$, $i(x_b) = i'(x_b) + i_b$ and $i'(x_p) = i''(x_p) + i_p$ \qquad (8b)

To solve these equations, we consider two cases, $x_p < x_b$ and $x_p > x_b$, separately. At first, differentiating equations (3), (4a), (5a) leads to three second order equations that have solutions in the following form,

$$i = C_1 e^{\tilde{B}x} + C_2 e^{-\tilde{B}x} \qquad (9)$$

$$i' = C_3 e^{\tilde{B}x} + C_4 e^{-\tilde{B}x} \qquad (10)$$

$$i'' = C_5 e^{\tilde{B}x} + C_6 e^{-\tilde{B}x} \qquad (11),$$

$$\text{where} \quad \tilde{B} = \sqrt{\frac{r_l}{r_t\left(\frac{1}{2} + f(B)\right)}}.$$

Inserting equations (9)-(11) into equations (3)-(8a) results in a set of linear equations. The constants, $C_1$-$C_6$ together with $V(x_p)$, can be determined by solving these equations numerically. The voltage at measurement point $x_p$ can then be obtained from equation 5(a) or 4(b).

## 5. Fiber parameters that affect the sensing performance

Fiber length, capacitance per unit length and resistivity of the conductive film are main parameters that affect the electrical properties of capacitor fibers. In this section we will study how these parameters affect the performance of the fiber as a touch sensor, and how we can control these parameters to make the fiber exhibits the best sensing properties.

5.1 The influence of fiber length

To study the effect of fiber length on its sensing performance, we take two pieces of fibers, named sample #1 and sample #2, from the same batch of drawing. Sample #1 with a length of 123mm is a part of sample #2 with a length of 246mm. Thus they share the same electrical and geometric parameters. The samples have an equal diameter around 1.1mm. The geometric parameters, such as the thickness and width of films ($d_c$, $d_i$ and $S$), can be measured from microscopic observations or calculated from the preform geometry and draw-down ratio. The fiber's electrical parameters were characterised independently by the method reported in reference [35]. The measured capacitance per unit length for both fibers equals 93nF/m. The volume resistivity of the conductive film, calculated from transverse resistance $r_t$, equals 4.5 Ωm. Figure 4 shows a comparison of the experimental data and model predictions for sample #1 and #2 at frequencies of 1k and 10k Hz. The data below 100Hz are not displayed because the disturbance caused by finger touch at these frequencies is very small.

Figure 4(a) is the voltage distribution on both fibers without finger touch. Symbols and curves represent experimental data and model predictions respectively. We can see that both fibers exhibit a voltage drop at the left end, which becomes more prominent as frequency goes higher. This phenomenon is precisely predicted by the model. In order to find the reasons that cause this observation, we made calculations by a model without the oscilloscope in the circuit, or simply set the value of $C_p$ extremely small and the value of $R_p$ extremely higher in the present model. The result shows that if we neglect the effect of the oscilloscope the voltage becomes a constant without any drop when the measurement point approaches the left end of the fiber. This implies that this voltage drop on the left end of the fiber is rather caused by the decreased impedance of the oscilloscope at higher frequencies than by the fiber itself. From figure 4(a) we can observe that both fibers have their voltage reduced to zero because the right end of the fiber is grounded. We can also see that the longer fiber, denoted by the blue color, has a longer part in the middle with its voltage be a constant, while the short fiber, denoted by the red color, has more part of its length with a changing voltage. This can be rationalized by the distributed nature of the fiber. The edge effect becomes more dominant when the fiber length decreases. In Figure 4(a) we compare the effect of frequency on the fiber performance. For both fibers the overall voltage decreases as frequency increases, which is related to the capacitance and resistance of the fiber.

Figure 4(b) shows the responses of the fiber caused by a finger touch at a fixed position of x≈0.3L. We can observe that the voltage has a reduction at the exact point of touch for both fibers. Longer fiber (blue color) exhibits a relatively deeper dip than the short fiber at the same frequencies. Both fibers have much larger reductions of voltage at touch point at 10Hz than at 1Hz.

Figure 4(c) displays the responses of the fiber when a finger moves on the fiber surface from the left end to the right end. The voltage at the left end $V(0)$, normalized by its untouched value $V_N(0)$, is plotted as a function of the touch position $x_b$. We can see that the shorter fiber have a relative much larger part of its length that responses to the finger touch than the longer one. This indicates that if we use the mapping of $V(0)$ versus $x_b$ to sense the touch position, the short fiber will performance better. We can also see that the both fibers become more sensitive at higher frequencies.

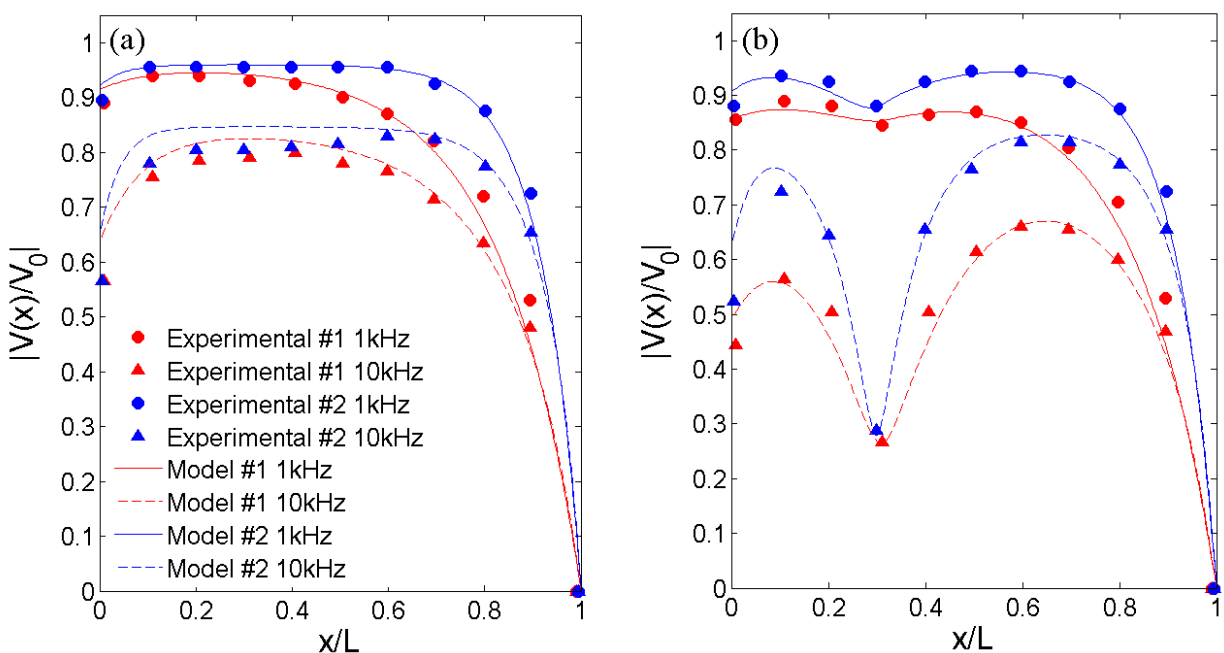

Figure 4 Comparison of the performance of fibers with different length. Sample #1, with a length of L=123mm, is denoted by red dots (1kHz) and triangles (10kHz). Sample #2, with a length of L=246mm, is denoted by blue dots (1kHz) and triangles (10kHz). The model predictions are represented by curves with the same color as the experimental data according to different samples. (a)The voltage distribution on the fiber surface without finger touch. (b) The voltage distribution on the fiber surface with at finger touch at a position of x≈0.3L.

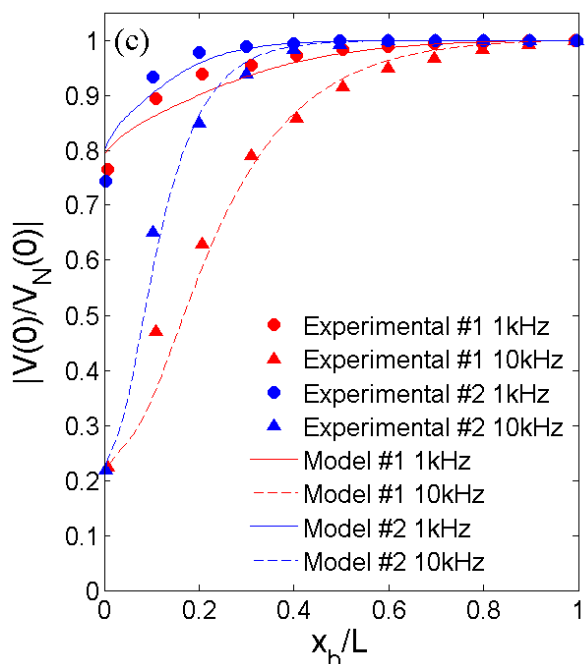

Figure 4 (c) Comparison of responses of the fiber when a finger moves on the surface of a fiber with different length. Sample #1, with a length of L=123mm, is denoted by red dots (1kHz) and triangles (10kHz). Sample #2, with a length of L=246mm, is denoted by blue dots (1kHz) and triangles (10kHz). The model predictions are represented by curves with the same color as the experimental data according to different samples.

2.1. The influence of capacitance per unit length

Capacitance per unit length, $C_F/L$, is one of the important parameters of capacitor fibers. It is independent on the diameter of fiber and drawing conditions but only dependent on the geometry of its preform [35]. To study the influence of the $C_F/L$ on fiber sensing, we select three pieces of fibers drawing from three performs with different geometric parameters. The samples displayed in figure 5 are selected among fibers with their electrical properties are carefully measured by the way presented in reference [35]. They have the same volume resistivity (2Ωm), same length (122mm) and similar diameters (1-1.2mm). The values of $C_F/L$ for sample #1, #2 and #3 are measured as 40, 65 and 95nF/m respectively. Although the sensing experiments of these samples were conducted at five frequencies of 1, 10, 100, 1k and 10Hz, only the data at 10 Hz is present here because the fibers have the most prominent responses at this frequency. The results calculated from the model are also displayed in the figure. Figure 5(a) displays the voltage distribution of three fibers without finger touch. The fiber with a smaller value of $C_F/L$ has a longer part that shows a constant voltage than the fiber with a larger value of $C_F/L$. This indicates that the distributed nature is more prominent for fibers with larger capacitance. Figure 5(b) exhibits the variation of voltage distribution in response to a fixed finger touch at x≈0.3L. Sample #2 of 65nF/m exhibits the deepest voltage dip at the touch point. Figure 5(c) shows how the voltage at the left end of fiber, V(0), varies as finger moves on the fiber surface. Sample #3 with the highest capacitance shows the best sensing performance as it has the largest part of this length that was responded to the touch among three samples. These results are identical to the model predictions.

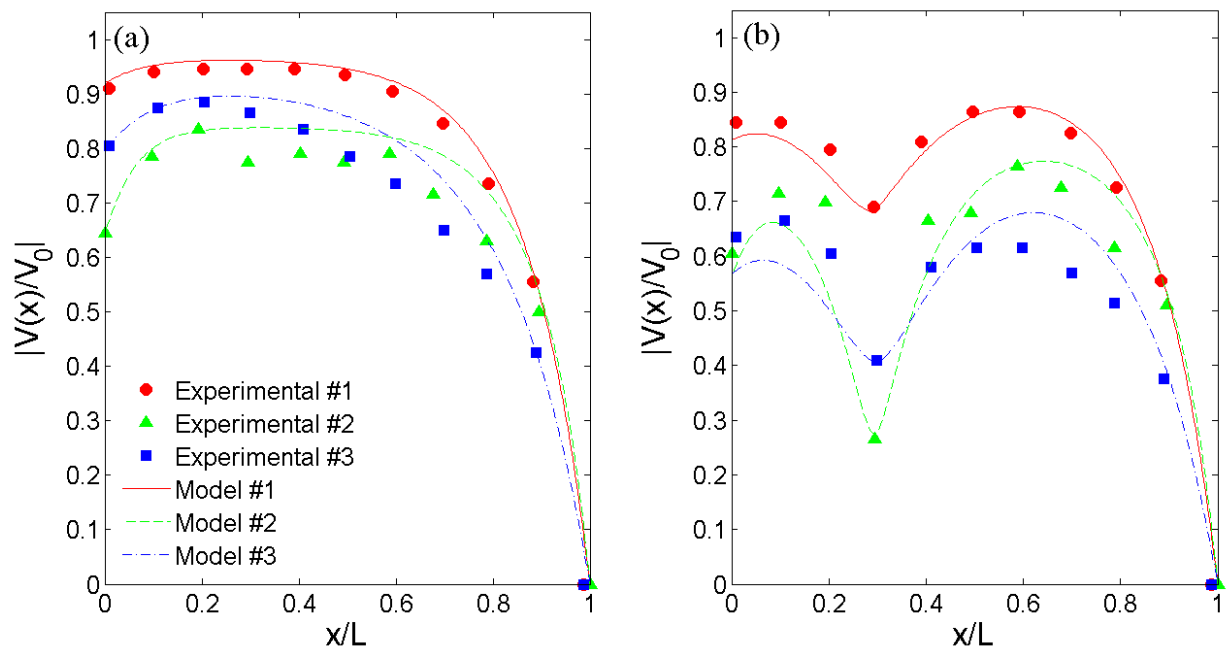

Figure 5. Comparison of the performance of fibers at 10kHz with different capacitance per unit length. Sample #1 with $C_F$=40nF/m is denoted by red dots; sample #2 with $C_F$=65nF/m is denoted by green triangles; Sample #3 with $C_F$=95nF/m is denoted by blue squares. (a)The voltage distribution on the fiber surface without finger touch. (b) The voltage distribution on the fiber surface with a finger touch at a position of x≈0.3L. The model predictions are represented by curves with the same color as the experimental data according to different samples.

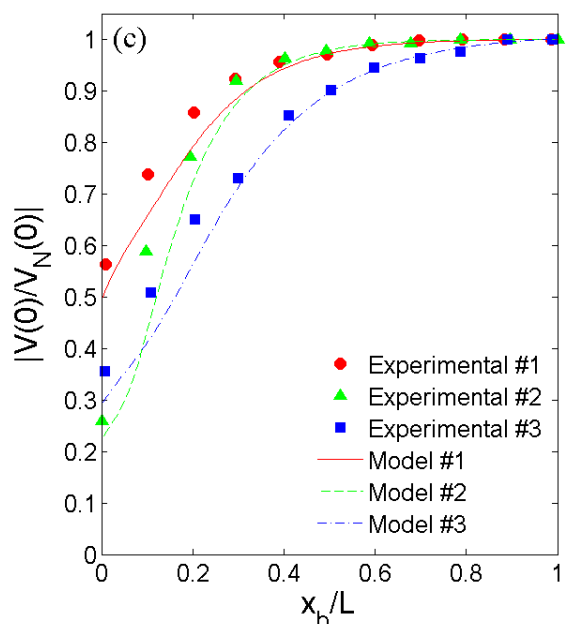

Figure 5 (c) The voltage at the left end of the fiber versus touch position on the surface of a fiber with different capacitance per unit length. Sample #1 with $C_F$=40nF/m is denoted by red dots; sample #2 with $C_F$=65nF/m is denoted by green triangles; Sample #3 with $C_F$=95nF/m is denoted by blue squares. The model predictions are represented by curves with the same color as the experimental data according to different samples.

2.2. The influence of resistivity

The volume resistivity of conductive layers, $\rho_v$, is another important parameter characterizing the capacitor fiber. It relates to the equivalent transverse resistance of the fiber by the way shown in equation (2). Generally a higher volume resistivity of the conductive layer leads to a higher equivalent capacitor resistance. To study the influence of $\rho_v$ on the sensing behaviour of fibers, we indentified three samples, namely #1, #2 and #3, with volume resistivities of 3 Ωm, 8 Ωm and 180Ωm, respectively. They share the same length of 120mm and the same capacitance per unit length of 95nF/m. Figure 6 displays their responses to finger touches at frequencies of 1k Hz (6(a) and 6(b)) and 10k Hz (6(c) and 6(d)). As a comparison, the model predictions based on aboved parameters are also presented. Figure 6(a) is the voltage distribution of the fiber at 1kHz when a finger touch is placed at positions closed to x=0.3L-0.4L. The figure indicates that as volume resistivity increase, the fiber becomes more sensitive to the finger touch and a deeper dip on voltage distribution can be observed. Figure 6(b) shows that the voltage measured at 1kHz at the left end of the fiber exhibits larger variations as finger moves on the fiber with a higher resistivity. Thus the fiber #3, with the highest resistivity, shows best performance as a touch sensor among three samples. However, at 10kHz the fiber with lower resistivity becomes more sensitive to the finger touch, as shown in figure 6(c). The performance of fibers with different resistivities also becomes very close, as presented in Figure 6(d). If we use the mapping between $|V(0)/V_N(0)|$ and $x_b$ to monitoring the finger touch, sample #2 with a moderate resistivity of 8Ωm exhibits the best sensing performance at 10k Hz. A further comparison of data at different frequencies as shown in figure 6(b) and 6(d) indicates that the best sensing performance was exhibited by sample #3 with the highest resistivity at the frequency of 1kHz. This indicates that by increasing the volume resistivity of fiber we can reduce the best performance frequency from 10kHz to 1kHz. This makes the application of the fiber sensor more feasible regarding the instrument requirement.

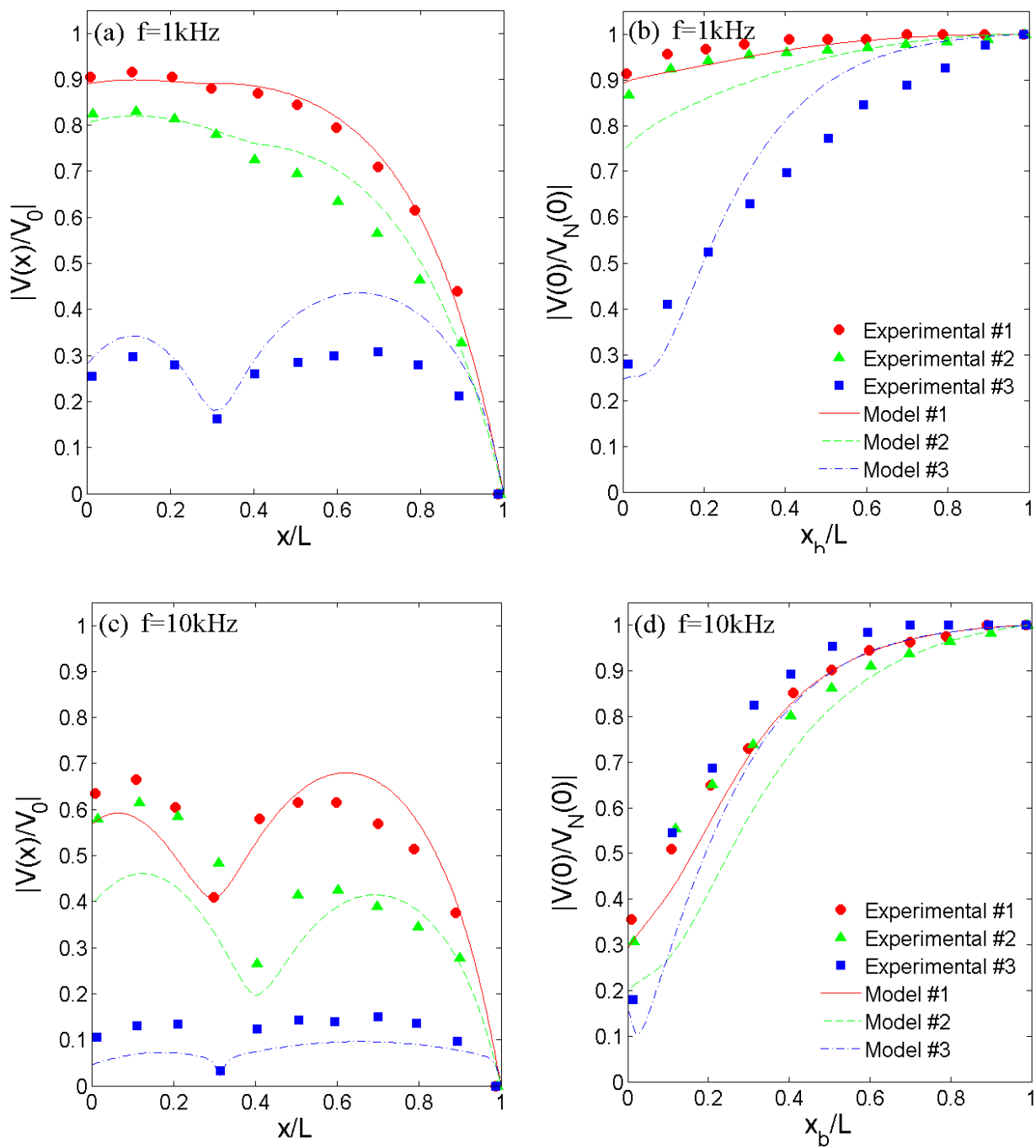

Figure 6. Comparison of the performance of fibers with different volume resistivity. Sample #1 with $\rho_v=3$ Ωm is denoted by red dots; sample #2 with $\rho_v=8$ Ωm is denoted by green triangles; Sample #3 with $\rho_v=180$ Ωm is denoted by blue squares. (a) The voltage distribution on the fiber surface with a finger touch at a position of $x\approx 0.3L$ at f=1kHz. (b) The voltage at the left end of the fiber versus touch position on the surface of a fiber at 1kHz. The model predictions are represented by curves with the same color as the experimental data according to different samples. (c) and (d) display the same experiments as (a) and (b) respectively, but at 10kHz.

To conclude, the above discussions on the effects of parameters show that a short fiber with larger capacitance per unit length and higher resistivity will perform better at frequency around 1kHz.

## 6  A textile sensor woven by capacitor fibers

So far we have discussed electrical properties of the individual capacitor fibers in terms of touch sensing. Flexibility and good overall mechanical properties of these fibers as well as their endurance towards deformations during the weaving process allow us to use them as building blocks of the touch sensing textile that we present in this section.

We used a Leclerc table loom to integrate 15 capacitor fibers into the wool-based textile. Resulting 15x10 cm textile and its equivalent scheme are presented at Fig. 7 (black lines correspond to the capacitor fibers, purple and green ones – to the wool base).

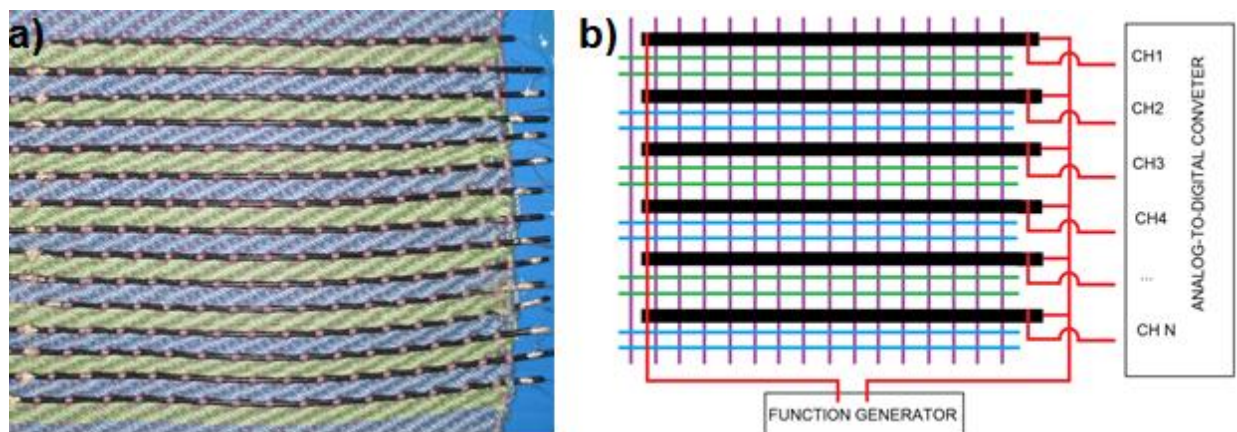

Figure 7. (a) Woven 2D sensing textile with integrated array of the capacitor fibers and b) its schematic representation. All the connections are done using a cooper wire (denoted by red lines) with a diameter of 0.2 mm. Fibers in the array have common ground and source, but are polled individually.

The inner electrode of each fiber is connected to the common source (function generator that provided a sinusoidal signal at 1 kHz with the amplitude of 4 V) and outer electrode (cladding) is connected to the multichannel Analog-to-Digital Convertor (National Instruments USB-6343 X-series DAQ). The connections are done using thin cooper wire (diameter 0.2 mm) and secured with a conductive epoxy. The ADC board is plugged into a PC and the custom software developed in our lab shows touch position on the computer screen (Fig. 8)

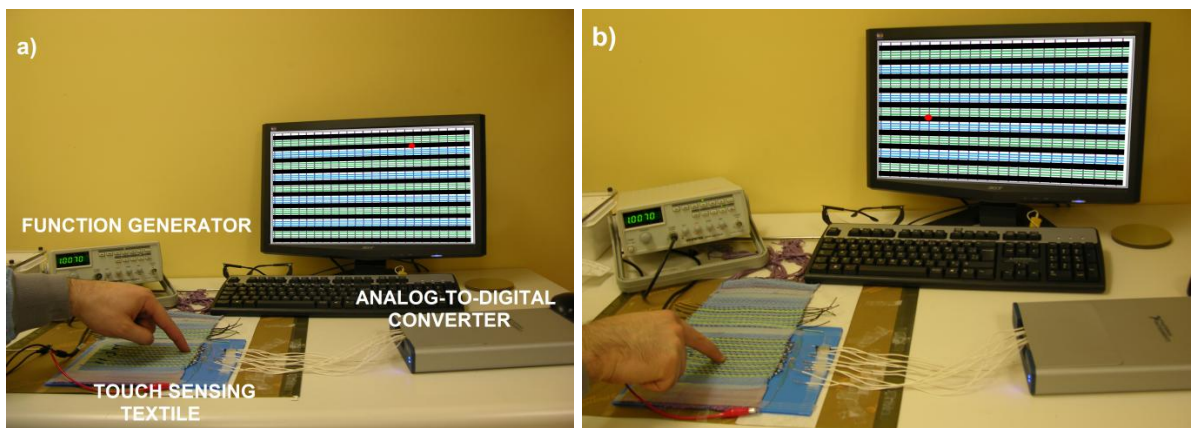

Figure 8. a) Experimental setup for the touch measurements of the sensing e-textile includes function generator (providing sinusoidal signal at 1 kHz with the amplitude of 4 V) and the ADC board connected to the PC. b) Touch position is shown on the computer screen.

6.1 Sensing principle

Since voltage response of the individual capacitor fiber depends on the position of the touching finger, it becomes possible to build so-called "map" for each fiber incorporated into the textile. An example of such "map" – voltage drop vs. touch position dependence - is shown at Fig 9 a). Here and after we count the distance from the point where the fiber is connected to the cooper wire. Thus by comparing measured signal with the "map" we can determine the touch position. And by polling each fiber individually we can create flexible 2D touch-sensor using only 1D array of capacitor fibers. From Fig. 9 a) one can notice that first, in general sensitivity (voltage drop) is higher closer to the beginning of the fiber and second, that the overall effective length of the fiber is around 15 cm. To our mind both these features could be explained by relatively high surface resistance of the outer cladding.

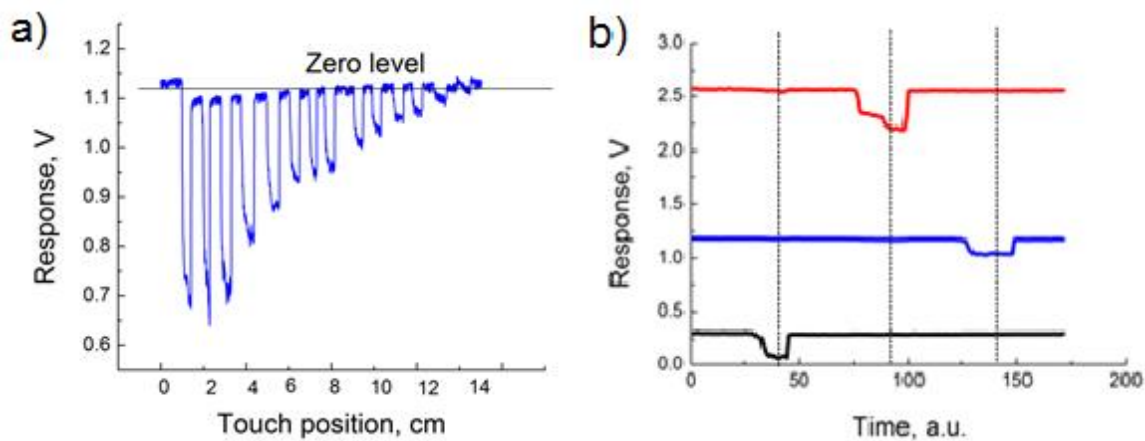

A fully woven touchpad sensor based on soft capacitor fibers, April 2011                    17Figure 9. a) A "map" (voltage drop vs. touch position dependence) of the capacitor fiber. b) Voltage response of the three fibers incorporated into the textile measured simultaneously demonstrates no sufficient cross-talk between them. Gaps on the graphs correspond to the touch.

6.2 Cross-talk and channel calibration.

For any touch sensor there are several important aspects that one needs to pay attention: whether it returns back to "idle" state after the touch, whether there is any cross-talk between the channels that may lead to the false response and whether any calibration for the particular user is needed.

As it could be seen from Fig. 9 that after the touch our textile sensor returns back to the stable "idle" level. Response from three different channels (Fig 9 b) was record simultaneously in order to answer the second question – possible cross-talk between the channels. And no significant influence at other channels could be seen when one of the fibers is touched. Different "idle" levels of the fibers at Fig. 9 b) are explained by their slightly different electrical properties (capacitance, resistance).

The third consideration is calibration for the particular user. As it was shown, touching the sensor with a finger is equivalent to adding to the circuit corresponding capacitance and resistance of a human body connected in series. Electrical parameters of the human body may vary from person to person depending on many factors, thus changing the absolute value of the touch response for the capacitor fiber ("fiber map"), increasing or decreasing gaps on the graph at Fig 9 a). Nevertheless overall principle will stay the same and the calibration procedure will involve only determination of the corresponding shift, which can be done by touching each fiber at zero-point. After that "fiber map" can be transformed within the software to match particular user.

# 7 Conclusion

In conclusion, we have demonstrated that soft capacitor fibers can be used as capacitive touch sensors and successfully integrated within a flexible textile medium. We studied the effects of finger touch on the voltage distribution of a single soft capacitor fiber, and discussed the influence of parameters such as fiber length, capacitance per unit length and resistivity of the conductive fiber on its sensing performance. A theoretical ladder network model was proposed to describe and accurately predict the electrical behavior of the device. The predictions yielded by this model were found to be in very good agreement with experimental results and thus allowed us to determine the optimal operating parameters of the fabricated two-dimensional touch-sensing textile prototype. The prototype demonstrated high sensitivity to finger interrogations across its surface, and showed good reliability against false responses.